\documentclass[aps,prb,twocolumn,groupedaddress,showpacs]{revtex4-1}
\usepackage{graphicx}
\usepackage[utf8]{inputenc}
\usepackage{tikz}
\usetikzlibrary{matrix,arrows,decorations.pathmorphing,shapes,snakes}
\usepackage{verbatim}
\usepackage{multirow}
\usetikzlibrary{shapes,snakes}
\usepackage{amsmath,amssymb}
\begin{document}

%\date{\today}
\title{Evidence of phonon-charge-density-waves coupling in ErTe$_3$}
\author{N. Lazarevi\'c and Z. V. Popovi\'c}
\affiliation{Center for Solid State Physics and New Materials, Institute of Physics, University of Belgrade,
Pregrevica 118, 11080 Belgrade, Serbia}
\author{$^{\ast }$Rongwei Hu and C. Petrovic}
\affiliation{Condensed Matter Physics and Materials Science Department, Brookhaven
National Laboratory, Upton, New York 11973-5000, USA}

\begin{abstract}
The vibrational properties of ErTe$_3$ were investigated using Raman spectroscopy and analyzed on the basis of peculiarities of the RTe$_3$ crystal structure. Four Raman active modes for the undistorted structure, predicted by factor-group analysis, are experimentally observed and assigned according to diperiodic symmetry of the ErTe$_3$ layer. By analyzing temperature dependence of the Raman mode energy and intensity we have provided the clear evidence that all Raman modes, active in the normal phase, are coupled to the charge density waves. In addition, new modes have been observed in the distorted state.
\end{abstract}

%\pacs{ 78.30.-j; 71.45.Lr; 63.20.K-; )
%\maketitle

\maketitle

\section{Introduction}
Low-dimensional systems with highly anisotropic electronic structure often exhibit electronic instabilities that can lead to the superconductivity or charge-density waves (CDW). CDW can be created through a Fermi surface (FS) nesting effect. If the FS can be nested with one q-vector of a particular phonon mode, the ground state energy can be reduced by electron-phonon coupling resulting in gaps opening at the FS and the creation of a new ground state with broken symmetry. FS for an ideal one dimensional (1D) system consists of two points, so it exhibits perfect nesting at $q=2k_f$. In real systems, a good nesting is reduced only to some parts of the FS. In this cases CDW gap is expected to open only on the best nested FS parts and the system can remain metallic in its new ground state.

In the recent years the high-temperature CDW family of rare-earth tritellurides (RTe$_3$) attracted a lot of attention.\cite{1} A wide range of tunable parameters in the RTe$_3$ offers unique opportunity for the study of the CDW formation.\cite{2,3,4,5,6,7} For the layered quasi-2D material with tetragonal symmetry, the CDW ground state can either be bidirectional (checkerboards) or unidirectional (stripes) with a reduced symmetry.\cite{8} The high resolution x-ray diffraction,\cite{5,6,7} angle resolved photoemission spectroscopy\cite{9,10} and femtosecond pump-probe spectroscopy\cite{11} showed that lighter rare-earth tritellurides (i.e. R=La-Tb) host unidirectional incommensurate CDW well above room temperature. For the heavy rare-earth tritellurides (i.e. R=Dy-Tm) transition temperature $T_{CDW_1}$ ($T_{CDW_1}$=265 K for ErTe$_3$ )resides below the room temperature with another transition to a bidirectional CDW state at $T_{CDW_2}$ ($T_{CDW_2}$=158 K for ErTe$_3$).\cite{7,11a}

The Raman scattering measurements of RTe$_3$ were analyzed, to the best of our knowledge, only in Refs. \onlinecite{15} and \onlinecite{18}. \emph{Lavagnini et al.}\cite{15} reported the Raman scattering spectra of RTe$_3$ (R=La, Ce, Pr, Nd, Sm, Gd, Dy) at ambient pressure and high pressure Raman spectra of LaTe$_3$ and CeTe$_3$. In the case of the room temperature  Raman spectra of LaTe$_3$, they observed five modes at about 72 cm$^{-1}$, 88 cm$^{-1}$, 98 cm$^{-1}$, 109 cm$^{-1}$ and 136 cm$^{-1}$ and labeled them as P1-P5. Polarization dependent measurements showed that the P4 peak at about 109 cm$^{-1}$ has 90$^{\circ}$ intensity change periodicity. They concluded that this mode can not be assigned within the undistorted structure and consequently assigned it as the B$_1$ symmetry mode of the distorted structure.\cite{15} In Ref. \onlinecite{18} amplitudon modes are observed and analyzed for DyTe$_{3}$ and LaTe$_{3}$.

In this work we have measured Raman scattering spectra of ErTe$_{3}$ in normal and in CDW state. Our polarization dependent Raman scattering measurements above $T_{CDW_1}$ showed that P4 mode\cite{15}, at about 120 cm$^{-1}$, persist in the normal state of ErTe$_{3}$. Appearance of this mode in the normal state and its B$_{1g}$ symmetry leads us to a conclusion  that the orthorhombic symmetry of the crystal does not play dominant role in the phonon properties of RTe$_{3}$. We have found that factor group analysis of the layer symmetry predicts 3A$_{1g}$+B$_{1g}$ modes to be observed in Raman scattering experiment from (010) plane of ErTe$_3$, in complete agreement with experimental findings. Temperature dependance of Raman mode energies  and intensities have shown strong coupling of phonons with the CDW gaps.

\section{Experiment}

Single crystals of ErTe$_3$ were grown using flux method.\cite{12b} The Raman scattering measurements were performed using Jobin Yvon T64000 Raman system in micro-Raman configuration. The 514.5 nm line of an Ar$^{+}$/Kr$^{+}$ mixed gas laser was used as an excitation source. All Raman scattering measurements presented in this work were performed from the (010) plane of ErTe$_{3}$ single crystals. Low temperature measurements were performed using KONTI CryoVac continuous Helium flow cryostat with 0.5 mm thick window. Focusing of the laser beam was realized with a long distance microscope objective (magnification $50\times $).  We have found that laser power level of 0.01 mW on the sample is sufficient to obtain Raman signal and, except signal to noise ratio, no changes of the spectra were observed as a consequence of laser heating by further lowering laser power. The corresponding excitation power density was less then 0.05 kW/cm$^{2}$.

\begin{figure}
\includegraphics[width=0.45\textwidth]{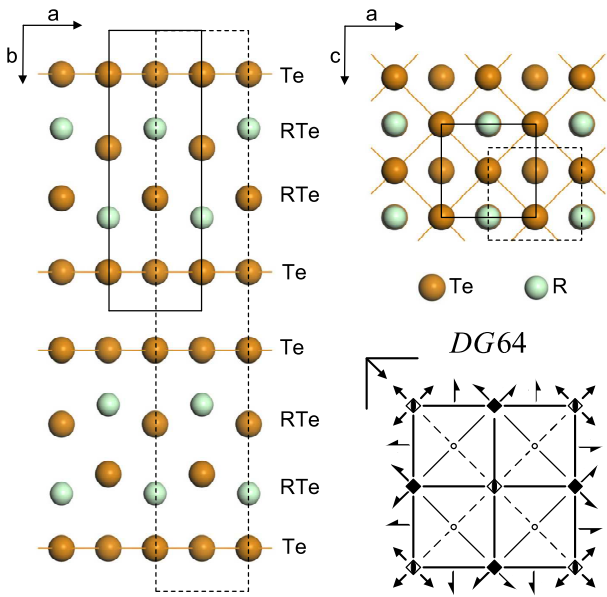}
\caption{Schematic representation of RTe$_3$ crystal structure. Dashed lines represent unit cell of RTe$_3$ crystal structure. Unit cell of single RTe$_3$ layer is represented with solid lines. Diagram (right bottom) represents symmetry operations of the DG 64 diperiodic symmetry group.\cite{14}}
\label{fig1}
\end{figure}

\section{Results and discussion}

ErTe$_3$ crystallizes in the $Cmcm$ (D$_{2h}^{17}$) space group orthorhombic structure with unit cell parameters $a=4.31$ {\AA}, $b=24.45$ {\AA}, $c=4.31$ {\AA} and $Z=4$.\cite{12} All ions are in the $c$ Wyckoff position (C$_{2v}$ site symmetry).\cite{12} The unit cell of ErTe$_3$ crystal structure (Fig.\ref{fig1}) is built up of corrugated double layers of ErTe sandwiched by planar square nets of Te making ErTe$_3$ slabs stacked along the b-axis. These slabs are mutually connected only by van der Waals forces.\cite{4}

Factor group analysis (FGA) for the crystal symmetry gives the normal-mode distribution, $\Gamma=4A_{g}+4B_{1g}+4B_{3g}+B_{1u}+B_{2u}+B_{3u}$. Based on the corresponding Raman tensors and the selection rules (see Table \ref{tab.1}), we found that only A$_{g}$ symmetry modes can be observed for our experimental configuration, in which both incident and scattered light are polarized parallel to the ($ac$) crystal plane.

%%%%%%%%%%%%%%%%%%%%%%%%%%%%%%%%%%%%%%%%%%%%%%%%%%%%%%%%%%%%%%%%%%%%%%%%%%%%%%%%%%%%%%%%

\tikzset{node style nd/.style ={text badly centered, midway,auto, minimum size=0.1cm, minimum width=2.9cm}}

\begin{table*}
\caption{Raman tensors and compatibility relations for the layer and the crystal symmetry of the RTe$_3$.}
\label{tab.1}\center
\begin{ruledtabular}
\begin{tabular}{c}

\begin{tikzpicture}

\matrix [nodes = {node style nd}, column sep=0.6cm, row sep=0.0cm]
{
\node[rounded corners, text width=3.5cm,minimum height=0.8cm,fill=gray!20] (1) {Raman tensors for tetragonal symmetry}; & \node[rounded corners, text width=3.5cm,minimum height=0.6cm,fill=gray!20] (2) {RTe$_3$ slab DG64}; & & \node[rounded corners, text width=3.5cm,minimum height=0.6cm,fill=gray!20] (3) {Crystal D$_{2h}$}; & \node[rounded corners, text width=3.5cm,minimum height=0.8cm,fill=gray!20] (4) {Raman tensors for orthorhombic symmetry};\\

 \node(a1) {\begin{minipage}{0.15\textwidth}
        A$_{1g}$=$
\begin{pmatrix}
A &  &  \\[-0.2cm]
 & A &  \\[-0.2cm]
 &  & B
\end{pmatrix}$
    \end{minipage}
}; & \node(c1) {3$A_{1g}(\alpha_{xx}+\alpha_{yy}, \alpha_{zz})$}; & &\node(d1) {4$A_{g}(\alpha_{xx}, \alpha_{yy}, \alpha_{zz})$};&\node(b1) {\begin{minipage}{0.15\textwidth}
        A$_{g}$=$
\begin{pmatrix}
A &  &  \\[-0.2cm]
 & B &  \\[-0.2cm]
 &  & C
\end{pmatrix}$
    \end{minipage}
};\\

 \node(a2) {\begin{minipage}{0.15\textwidth}
B$_{1g}$=$
\begin{pmatrix}
C & &  \\[-0.2cm]
 &-C&  \\[-0.2cm]
 &  &
\end{pmatrix}
 $
    \end{minipage}
  };& \node(c6) {$B_{1g}(\alpha_{xx}-\alpha_{yy})$}; & &\node(d2) {4$B_{1g}(\alpha_{xy})$};& \node(b2) {\begin{minipage}{0.15\textwidth}
B$_{1g}$=$
\begin{pmatrix}
 & D &  \\[-0.2cm]
D & &  \\[-0.2cm]
 & &
\end{pmatrix}
 $
    \end{minipage}
};\\

\node(a3) {\begin{minipage}{0.15\textwidth}
B$_{2g}$=$
\begin{pmatrix}
 & D &  \\[-0.2cm]
D &  &  \\[-0.2cm]
 &  &
\end{pmatrix}
 $
    \end{minipage}
};& \node(c2) {3$A_{2u}(E\parallel z)$}; & &\node(d3) {4$B_{1u}(E\parallel z)$}; &\node(b3) {\begin{minipage}{0.15\textwidth}
B$_{2g}$=$
\begin{pmatrix}
 & & E \\[-0.2cm]
 & &  \\[-0.2cm]
E &  &
\end{pmatrix}
 $
    \end{minipage}
};\\

\node(a4) {\begin{minipage}{0.15\textwidth}
       E$_{g}$=$
\begin{pmatrix}
 &  &-E \\[-0.2cm]
 &  &  \\[-0.2cm]
-E& &
\end{pmatrix}$
    \end{minipage}
}; & \node(c5) {$B_{2u}(silent)$};  & &\node(d4) {4$B_{2u}(E\parallel y)$};& \node(b4) {\begin{minipage}{0.15\textwidth}
       B$_{3g}$=$
\begin{pmatrix}
 & & \\[-0.2cm]
 & & F \\[-0.2cm]
 & F &
\end{pmatrix}$
    \end{minipage}
};\\

  \node(a5) {\begin{minipage}{0.15\textwidth}
       E$_{g}$=$
\begin{pmatrix}
 &  &  \\[-0.2cm]
 &  & E \\[-0.2cm]
 & E &
\end{pmatrix}
 $
    \end{minipage}
};& \node(c3) {4$E_{g}(\alpha_{xz}, \alpha_{yz})$}; & &\node(d5) {4$B_{3g}(\alpha_{yz})$};& \\[0.0cm]

\node(b6) {\begin{minipage}{0.20\textwidth}
       $x=a$\\
       $y=c$\\
       $z=b$\\
    \end{minipage} }; & \node(c4) {4$E_{u}(E\parallel x, E\parallel y)$}; & &\node(d6) {4$B_{3u}(E\parallel x)$};&\node(b5) {\begin{minipage}{0.20\textwidth}
       $x=a$\\
       $y=b$\\
       $z=c$\\
    \end{minipage} };\\
 };

%\draw [-] (1.south west) -- (4.south east);

\draw [-] (c6.south west) -- (c6.south east);
\draw [-] (c1.south west) -- (c1.south east);
\draw [-] (c2.south west) -- (c2.south east);
\draw [-] (c3.south west) -- (c3.south east);
\draw [-] (c4.south west) -- (c4.south east);
\draw [-] (c5.south west) -- (c5.south east);

\draw [-] (d1.south west) -- (d1.south east);
\draw [-] (d2.south west) -- (d2.south east);
\draw [-] (d3.south west) -- (d3.south east);
\draw [-] (d4.south west) -- (d4.south east);
\draw [-] (d5.south west) -- (d5.south east);
\draw [-] (d6.south west) -- (d6.south east);

\draw [dashed,-latex] (c1.south east) -- (d1.south west);
\draw [dashed,-latex] (c6.south east) -- (d1.south west);
\draw [dashed,-latex] (c2.south east) -- (d4.south west);
\draw [dashed,-latex] (c3.south east) -- (d5.south west);
\draw [dashed,-latex] (c3.south east) -- (d2.south west);
\draw [dashed,-latex] (c4.south east) -- (d6.south west);
\draw [dashed,-latex] (c4.south east) -- (d3.south west);
\draw [dashed,-latex] (c5.south east) -- (d4.south west);
\end{tikzpicture}

\end{tabular}
\end{ruledtabular}
\end{table*}

%%%%%%%%%%%%%%%%%%%%%%%%%%%%%%%%%%%%%%%%%%%%%%%%%%%%%%%%%%%%%%%%%%%%%%%%%%%%%%%%%%%%%%%%%%%%%%%%

\begin{figure}
\includegraphics[width=0.35\textwidth]{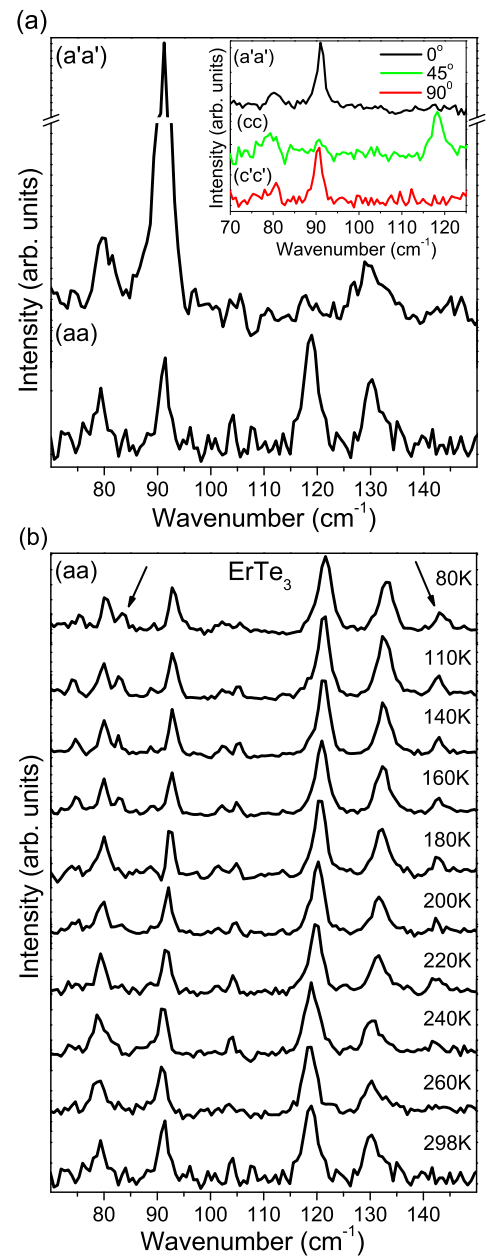}
\caption{(a) Raman scattering spectra of ErTe$_3$ single crystals measured at room temperature in parallel polarization for different sample orientations, where $a=[100]$, $c=[001]$, $a'=\frac{1}{\sqrt{2}}[101]$ and $c'=\frac{1}{\sqrt{2}}[10\bar{1}]$.  Inset: Raman scattering spectra of ErTe$_3$ measured for three different sample orientations. (b) Raman scattering spectra of ErTe$_3$ measured at different temperatures. New modes in the CDW state are denoted with arrows.}
\label{fig2}
\end{figure}

Fig. \ref{fig2} (a) shows room temperature Raman scattering spectra of ErTe$_3$ (normal state) measured in parallel polarization for different orientations of the sample. By changing an angle between incident light polarization and the crystal axes (by rotating the sample) one can find the position at which the Raman mode at about 120 cm$^{-1}$ disappears. The disappearance of this mode is shown to be periodical with the period of 90$^\circ$ (see inset in Fig. \ref{fig2} (a)). The same type of behavior was observed for the P4 mode in Ref. \onlinecite{15} for LaTe$_3$ in the CDW state. Since periodicity of the A$_{g}$ mode intensity change for the orthorhombic symmetry is 180$^\circ$, the P4 mode can not be assigned within the orthorhombic crystal symmetry. Because the crystal structure of RTe$_3$ consists of two RTe$_3$ layers mutually connected by weak van der Waals forces, it is natural to expected that layer symmetry dominates in vibration properties of RTe$_3$.

Layered crystals can be properly treated as molecular crystals in which the molecular unit is a sheet extended infinitely in two dimensions.\cite{13} Considering the vibrational properties of the individual layers, we break the periodicity in the direction perpendicular to the layer. Thus, for considering symmetry related properties of the layer, we have to use the diperiodic groups in three dimensions.\cite{14}

The structure of the ErTe$_3$ slab is shown in Fig. \ref{fig1}. The diperiodic symmetry group which fully describes symmetry of the ErTe$_3$ layer is DG 64, with two formula units (Z=2) per unit cell. Two R and two Te atoms are in $c$ (C$_{4v}$ site symmetry) and four Te atoms are in $f$ (C$_{2v}$ site symmetry) Wyckoff positions. FGA for the ErTe$_{3}$ slab gives, $\Gamma=3A_{1g}+B_{1g}+3A_{2u}+B_{2u}+4E_g+4E_u$. There are four Raman active modes ($3A_{1g}+B_{1g}$) to be observed in (010) plane. Correlation diagram between crystal (orthorhombic) and layer (tetragonal) symmetries is presented in the Table \ref{tab.1}. According to the Raman tensors for tetragonal symmetry (Table \ref{tab.1}), the B$_{1g}$ symmetry mode has intensity change periodicity of 90$^\circ$. From the selection rules, we have found that the B$_{1g}$ symmetry mode is Raman active (in parallel polarization configuration) for all orientations of the sample except $a'a'$ and $c'c'$. This is in agreement with our experimental findings. In this way we can assign P4 mode of ErTe$_3$ (peak around 120 cm$^{-1}$) as the B$_{1g}$ symmetry mode of the ErTe$_3$ slab. This means that 90$^\circ$ periodicity change of P4 mode intensity is not related with CDW state, as proposed in Ref. \onlinecite{15}. It is a consequence of the dominant role of ErTe$_3$ layers in the optical properties of the crystal.  The peaks at 132 cm$^{-1}$, 92 cm$^{-1}$ and 80 cm$^{-1}$ can be assigned as the A$_{1g}$ symmetry modes of the ErTe$_3$ slab.

We cannot exclude the twining effect as a cause of tetragonal symmetry dominance in Raman spectra of ErTe$_3$. Namely, due to the weak inter-layer van de Waals bonding, $b$-axis stacking faults can appear, resulting in twin formation. Such single crystals often can have regions for which the $a$ and $c$ axes are exchanged and sample exhibits an average 4-fold symmetry, even though it really comprises a superposition of two orthogonal orthorhombic symmetries.

ErTe$_3$ exhibits two successive second order phase transitions upon cooling at $T_{CDW_1}$=265 K and $T_{CDW_2}$=158 K,\cite{11a} followed by the modulation of electronic charges and unidirectional and bidirectional ground state formation.\cite{7} According to the Landau's theory of second order phase transitions, for each second order phase transition a characteristic order parameter can be introduced. In the case of CDW formation, order parameter is the CDW gap. Temperature dependance of the CDW gap can be written as $\Delta(T)/\Delta(0)\sim(1-T/T_{CDW})^{\beta}$, where $\beta$ is the critical exponent.

\begin{figure}
\includegraphics[width=0.45\textwidth]{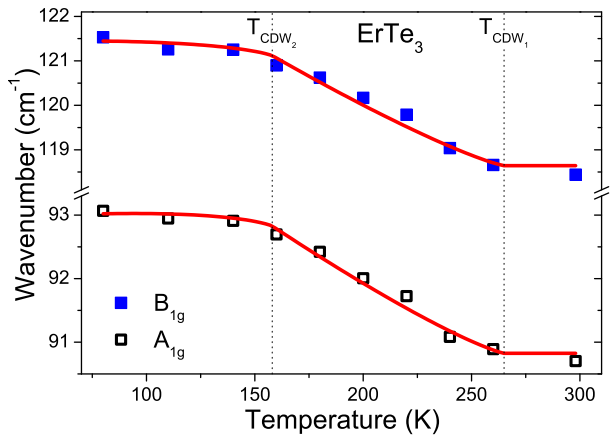}
\caption{Energy of the Raman active phonons of ErTe$_3$ for the undistorted structure as a function of temperature. Solid lines represent calculated spectra by using Eq. (\ref{eq2}). }
\label{fig3}
\end{figure}

Temperature dependances of the energy of the 92 cm$^{-1}$ (A$_{1g}$) and 120 cm$^{-1}$ (B$_{1g}$) modes of ErTe$_3$ are shown in Fig. \ref{fig3}. The temperature-induced anharmonicity effect\cite{16,17} can not describe changes of slope at $T_{CDW_1}$ and $T_{CDW_2}$. Thus, we must assume that the changes in energies come from charges redistribution in the second order phase transition. According to theoretical predictions for the Raman modes, active in the normal phase, the energy change induced by second order phase transition is proportional to the square of order parameter (CDW gap),\cite{20} $\Delta\omega(T)\sim\Delta(T)^2$. Having this in mind, for the energy shift induced by the second order phase transition, for $T<T_{CDW_i}$, we obtain:
 \begin{equation}\label{eq2a}
  \Delta\omega(T)_{CDW_i}=A_i(1-T/T_{CDW_i})^{2\beta},
\end{equation}
where $A_i$ is a coupling constant. If we neglect the interaction between two CDWs\cite{10} we can approximate the total change of the mode energy, Raman active in the "mother" phase (normal state), with:
\begin{equation}\label{eq2}
    \omega(T)=\omega_0+\Delta\omega_{CDW_1}+\Delta\omega_{CDW_2},
\end{equation}
where $\omega_0$ is a constant. Solid lines in Fig. \ref{fig3} represent the calculated spectra using Eq. (\ref{eq2}). An excellent agreement with the experimental data has been archived for $\beta=0.30(2)$, which is also in a complete agreement with the results ($\beta=0.3$) obtained in Refs. \onlinecite{11} and \onlinecite{18}. This is not surprising since amplitudon mode is nothing but the transverse fluctuation of the order parameter. This provides clear evidence for the tight coupling between the CDW gap and the Raman active modes.

\begin{figure}
\includegraphics[width=0.45\textwidth]{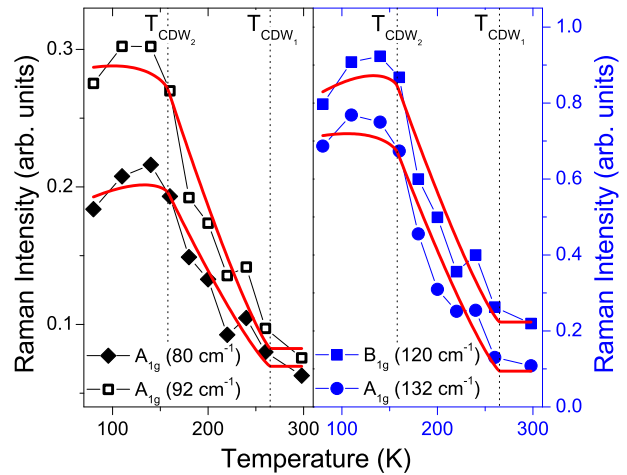}
\caption{Integrated intensity of the Raman active phonons of ErTe$_3$ for the undistorted structure as a function of temperature. Solid lines represent calculated spectra by using Eq. (\ref{eq4}). }
\label{fig4}
\end{figure}

Fig. \ref{fig4} shows temperature dependance of the intensity of the four modes, Raman active for the undistorted structure. One can see that all four modes exhibit the same type of temperature dependence of the intensity. According to Ref. \onlinecite{20}, change of intensity of the Raman mode upon the second order phase transition for the non-degenerate modes, Raman active in mother phase, is proportional to the square of order parameter, $\Delta I(T)\sim\Delta(T)^2$. For $T<T_{CDW_i}$,  we obtain:
 \begin{equation}\label{eq4a}
  \Delta I(T)_{CDW_i}=B_i(1-T/T_{CDW_i})^{2\beta},
\end{equation}
where $B_i$ is a coupling constant. By neglecting interaction between CDWs\cite{10}, we can approximate the Raman intensity temperature change of the modes, Raman active in the mother phase, with:
\begin{equation}\label{eq4}
    I(T)=I_0+\Delta I_{CDW_1}+\Delta I_{CDW_2},
\end{equation}
where $I_0$ is a constant. The spectra calculated using Eq. (\ref{eq4}) for $\beta=0.3$ (solid lines in Fig. (\ref{fig4})) scales rather well with the experimental data. This gives us additional evidence of coupling between Raman modes  and the CDW gaps.

Formation of the CDW state is followed by the development of a modulation in the density of the electronic charges which do not transform according
to the symmetry group that describes the ionic positions. Due to the Brillouin-zone folding and lowering the symmetry, new modes became observable in the distorted state. Because of the nature of the second order phase transition, one does not expect a sudden appearance or disappearance of phonons. In the vicinity of the phase transition, these modes additionally lose intensity and broaden due to fluctuations of order parameter. In the CDW state of the ErTe$_3$, we have observed additional modes (denoted with arrows  in Fig. \ref{fig2} (b)) at 143 cm$^{-1}$ and 83 cm$^{-1}$, which can not be assigned within the undistorted structure. We concluded that these modes are the consequence of the distortions. Although, theory predicts large number of such modes, many of them can not be observed due to the mode overlapping and their weak intensity.

\section{Conclusion}
The vibrational properties of ErTe$_3$ were investigated using Raman spectroscopy and analyzed on the basis of the RTe$_3$ crystal structure. We have concluded that vibrations of the RTe$_3$ layers play dominant role in the optical properties of the crystal. Four Raman active phonons predicted by FGA of the layer symmetry were observed and assigned. Temperature dependence of the energies and intensities of the modes, Raman active for the undistorted structure, have clearly shown coupling of phonons to the CDWs. Theoretical calculations perfectly map experimental data for the critical exponent $\beta=0.3$. In the distorted state of ErTe$_3$ additional Raman modes have been observed.
\section*{Acknowledgment}

This work was supported by the Serbian Ministry of Science and Technological Development under the Projects No. 141047,
No. ON171032, and No. III45018. Part of this work was carried out at the Brookhaven National Laboratory which is operated for the Office of Basic Energy Sciences, U.S. Department of Energy by Brookhaven Science Associates (DE-Ac02-98CH10886).

$^{\ast }$ Present address Ames Laboratory and Department of Physics and
Astronomy, Iowa State University, Ames, Iowa 50011, USA

\end{document}